\documentclass[final,3p,twocolumn,sort&compress]{elsarticle}

\usepackage{amsmath}
\usepackage{amssymb}
\usepackage[figurename=Fig.,labelsep=period]{caption}

\journal{Physics Letters B}

\begin{document}

\begin{frontmatter}




\title{A two-dimensional soliton system in the Maxwell-Chern-Simons \\
       gauge model}

\author[tpu]{A.Yu.~Loginov\corref{cor1}}
\ead{aloginov@tpu.ru}
\cortext[cor1]{Corresponding author}
\author[tpu]{V.V.~Gauzshtein}

\address[tpu]{Tomsk Polytechnic University, 634050 Tomsk, Russia}

\begin{abstract}
The $(2 + 1)$-dimensional Maxwell-Chern-Simons  gauge  model  consisting of two
complex scalar  fields  interacting  through  a  common  Abelian gauge field is
considered.
It is shown that  the  model  has  a  solution  that describes a soliton system
consisting of vortex and Q-ball constituents.
This two-dimensional soliton system possesses  a  quantized magnetic flux and a
quantized electric charge.
Moreover, the soliton system has a nonzero angular momentum.
Properties  of  this  vortex-Q-ball  system  are investigated by analytical and
numerical methods.
It is found that the  system   combines   properties  of a vortex and a Q-ball.
\end{abstract}


\end{frontmatter}

\section{Introduction}
\label{seq:I}

Topological solitons of $(2+1)$-dimensional field models play an important role
in various areas of field  theory,  physics of condensed  state, cosmology, and
hydrodynamics.
Among  them,  we  should  first  mention  vortices  of  the effective theory of
superconductivity \cite{abr}, vortices of the $(2+1)$-dimensional Abelian Higgs
model  \cite{nielsen},  and  lumps of  the $(2+1)$-dimensional nonlinear $O(3)$
sigma model \cite{belpol}.

In contrast  to  the  $(3 + 1)$-dimensional case, electrically charged solitons
do not exist in  the $(2 + 1)$-dimensional Maxwell electrodynamics for a fairly
straightforward reason: the  electric  field  goes  like $1/r$, so the electric
field's energy diverges logarithmically.
In $(2+1)$ dimensions, however, the dynamics of gauge field may be governed not
only by the Maxwell term but also by the Chern-Simons term \cite{JT, schonfeld,
DJT}.
In the presence of the  Chern-Simons term, a gauge field  becomes topologically
massive,  thus  making  possible  the existence of two-dimensional electrically
charged solitons.
These solitons exist  both  in   the  Maxwell-Chern-Simons  models  \cite{paul,
khare_rao_227, khare_255} and in the Chern-Simons models \cite{hong, jw1,  jw2,
bazeia_1991,  ghosh}   and    can    be   both  topological and nontopological.
Topological solitons are electrically charged  vortices, whereas nontopological
ones  are  two-dimensional  electrically  charged  spinning  (possessing an
angular momentum) Q-balls.
The numerical research of such  a  two-dimensional Q-ball has been performed in
\cite{deshaies_2006}.
The three-dimensional  counterparts  of  these Q-balls  have  been described in
\cite{volkov_2002, radu}.
More recently, the influence  of  the Chern-Simons term on electrically charged
and spinning solitons of several $(2 + 1)$-dimensional Abelian gauge models has
been studied in \cite{navarro_2017}.

In  this  Letter a  two-dimensional soliton system in  the Maxwell-Chern-Simons
gauge model is considered.
As well as in  the  Maxwell  gauge  model  \cite{loginov_plb_777}, the  soliton
system consists of a vortex  and  a Q-ball interacting through a common Abelian
gauge field.
This vortex-Q-ball system possesses a radial electric field, carries a quantized
magnetic flux,  and   has  a  nonzero  angular  momentum,  but  in  contrast to
\cite{loginov_plb_777}, it also has a quantized electric charge.
It is  shown  that  the vortex-Q-ball system combines properties of topological
and nontopological solitons.

\section{Lagrangian and field equations of the model}
\label{seq:II}

The Lagrangian density of the model is
\begin{eqnarray}
\mathcal{L} & = & -\frac{1}{4}F_{\mu \nu }F^{\mu \nu } + \frac{\mu }{4}\epsilon
^{\rho \sigma \tau}F_{\rho \sigma}A_{\tau} \nonumber \\
&& + \left( D_{\mu }\phi \right) ^{\ast }D^{\mu }\phi - V\left( \left\vert \phi
\right\vert \right) \nonumber \\
&& + \left( D_{\mu }\chi \right) ^{\ast }D^{\mu }\chi - U\left( \left\vert \chi
\right\vert \right),                                                  \label{1}
\end{eqnarray}
where  the  complex  scalar fields $\phi$ and $\chi$  are  minimally coupled to
the Abelian gauge field $A_{\mu}$ through the covariant derivatives:
\begin{equation}
D_{\mu }\phi =\partial_{\mu }\phi + ieA_{\mu }\phi,\quad
D_{\mu }\chi =\partial_{\mu }\chi + iqA_{\mu }\chi.                   \label{2}
\end{equation}
The self-interaction potentials used in this paper are the  same  as those used
in \cite{loginov_plb_777}:
\begin{align}
V\left( \left\vert \phi \right\vert \right)  &=\frac{\lambda }{2}\left(
\phi ^{\ast }\phi -v^{2}\right) ^{2}\!, \nonumber \\
U\left( \left\vert \chi \right\vert \right)  &=m^{2}\chi ^{\ast }\chi
-g\left( \chi^{\ast }\chi \right)^{2}+
h \left( \chi^{\ast }\chi \right)^{3}\!.                              \label{3}
\end{align}
In Eq.~(\ref{3}), $\lambda$, $g$, and $h$  are  the  positive  self-interaction
constants, $m$ is the mass of the scalar $\chi$-particle, and $v$ is the vacuum
average of the complex scalar field $\phi$.
We  suppose  that  the  parameters  $m$,  $g$,  and  $h$  satisfy the condition
\begin{equation}
\frac{g^{2}}{4m^{2}}<h<\frac{g^{2}}{3m^{2}}.                          \label{4}
\end{equation}
In this case, the  potential $U\left(\left\vert\chi\right\vert\right)$  has the
two minima: the  global  minimum  at $\chi = 0$ and a local one at some nonzero
$\left\vert \chi \right\vert$.

The model's action  $S = \int\mathcal{L}d^{3}x$  is  invariant  under the local
gauge transformations:
\begin{eqnarray}
\phi \left( x\right)  &\rightarrow &\phi ^{\prime }\left( x\right) = \exp
\left(-ie\Lambda \left( x\right) \right) \phi\left( x\right) , \nonumber \\
\chi \left( x\right)  &\rightarrow &\chi ^{\prime }\left( x\right) = \exp
\left(-iq\Lambda \left( x\right) \right) \chi\left( x\right) , \nonumber \\
A_{\mu }\left( x\right)  &\rightarrow &A_{\mu }^{\prime }\left( x\right)
=A_{\mu }\left( x\right) +\partial _{\mu }\Lambda \left( x\right)     \label{5}
\end{eqnarray}
if  the  local  gauge  parameter $\Lambda \left( x\right)$ decreases rapidly at
infinity.
Because of neutrality  of  the  Abelian  gauge  field $A_{\mu}$, the Lagrangian
density  (\ref{1}) is  also invariant  under  the  two independent global gauge
transformations:
\begin{eqnarray}
\phi \left( x\right)  &\rightarrow &\phi ^{\prime }\left( x\right) = \exp
\left(-i\alpha \right) \phi \left( x\right) , \nonumber \\
\chi \left( x\right)  &\rightarrow &\chi ^{\prime }\left( x\right) = \exp
\left(-i\beta \right) \chi \left( x\right).                           \label{6}
\end{eqnarray}
This  invariance leads to the two Noether currents:
\begin{eqnarray}
j_{\phi }^{\mu } &=& i\left[ \phi ^{\ast }D^{\mu }\phi -\left( D^{\mu }\phi
\right) ^{\ast }\phi \right], \nonumber \\
j_{\chi }^{\mu } &=& i\left[ \chi ^{\ast }D^{\mu }\chi -\left( D^{\mu }\chi
\right) ^{\ast }\chi \right].                                         \label{7}
\end{eqnarray}
Under the discrete  transformations  $C$,  $P$,  and $T$, the Chern-Simons term
$\mathcal{L}_{C\!S} = \mu \epsilon^{\rho\sigma\tau }F_{\rho \sigma}A_{\tau }/4$
behaves as follows:
\begin{equation}
\mathcal{L}_{\mathrm{CS}}^{\left( C\right) }=\mathcal{L}_{\mathrm{CS}},\;
\mathcal{L}_{\mathrm{CS}}^{\left( P\right) }=-\mathcal{L}_{\mathrm{CS}},\;
\mathcal{L}_{\mathrm{CS}}^{\left( T\right) }=-\mathcal{L}_{\mathrm{CS}}.
                                                                     \label{7a}
\end{equation}
It follows from Eq.~(\ref{7a}) that  the  Chern-Simons  term  breaks  the  $P$,
$CP$, and $T$-invariance of the model's Lagrangian.

The field equations of the model have the form:
\begin{flalign}
&\partial_{\mu }F^{\mu \nu }+\mu \left.\!\hspace{-0.05cm}\right.^{\ast }\!
\hspace{-0.05cm}F^{\nu }=j^{\nu },                                   \label{8a}
\\
& D_{\mu }D^{\mu }\phi +\lambda \left( \phi ^{\ast }\phi - v^{2}\right)
\phi = 0, \label{8b} \\
& D_{\mu }D^{\mu }\chi +\left( m^{2}-2g\left( \chi^{\ast }\chi \right)
+3h\left( \chi^{\ast }\chi \right)^{2}\right) \chi = 0,              \label{8c}
\end{flalign}
where the  dual  field  strength  $\left.\! \hspace{-0.05cm} \right.^{\ast } \!
\hspace{-0.05cm}F^{\nu} = \epsilon^{\nu \alpha \beta } F_{\alpha \beta}/2$, and
the electromagnetic  current  $j^{\nu}$  is  expressed in  terms of the Noether
currents:
\begin{equation}
j^{\nu } = e j_{\phi }^{\nu } + q j_{\chi }^{\nu }.                   \label{9}
\end{equation}
Integrating the left and right  hand  sides of  Eq.~(\ref{8a})  with  the index
$\nu = 0$ over the spatial plane, we  obtain  an important relation between the
electric charge and the magnetic flux:
\begin{equation}
Q = e Q_{\phi} + q Q_{\chi} = - \mu \Phi,                            \label{9b}
\end{equation}
where $Q_{\phi}=\int j_{\phi}^{0}d^{2}x$ and $Q_{\chi}=\int j_{\chi}^{0}d^{2}x$
are the conserved Noether charges.

The symmetric  energy-momentum  tensor  of  the  model  and  the  corresponding
expression for the energy density are written as: 
\begin{align}
T_{\mu \nu } =&-F_{\mu \lambda }F_{\nu }^{\;\lambda }+\frac{1}{4}g_{\mu
\nu }F_{\lambda \rho }F^{\lambda \rho } \nonumber \\
&+\left( D_{\mu }\phi \right) ^{\ast }D_{\nu }\phi +\left( D_{\nu }\phi
\right) ^{\ast }D_{\mu }\phi  \nonumber \\
&-g_{\mu \nu }\left( \left( D_{\mu }\phi \right) ^{\ast }D^{\mu }\phi
-V\left( \left\vert \phi \right\vert \right) \right) \nonumber \\
&+\left( D_{\mu }\chi \right) ^{\ast }D_{\nu }\chi +\left( D_{\nu }\chi
\right) ^{\ast }D_{\mu }\chi  \nonumber \\
&-g_{\mu \nu }\left( \left( D_{\mu }\chi \right) ^{\ast }D^{\mu }\chi
-U\left( \left\vert \chi \right\vert \right) \right),                \label{10}
\end{align}
\begin{align}
T_{00} = & \frac{1}{2}E_{i}E_{i}+\frac{1}{2}B^{2}              \label{11} \\
& +\left( D_{0}\phi \right) ^{\ast }D_{0}\phi + \left( D_{i}\phi \right)
^{\ast }D_{i}\phi +V\left( \left\vert \phi \right\vert \right) \nonumber  \\
& +\left( D_{0}\chi \right) ^{\ast }D_{0}\chi +\left( D_{i}\chi \right)
^{\ast }D_{i}\chi +U\left( \left\vert \chi \right\vert \right), \nonumber
\end{align}
where $E_{i} = F_{0i}$  are  the  components  of  electric  field  strength and
$B = -F_{12}$ is the magnetic field strength.

In this  paper we adopt the following gauge condition: $\partial _{0}\phi = 0$.
Using the analogy of Q-ball, we find a soliton solution of model (\ref{1}) that
minimizes  the  energy  $E = \int T_{00}d^{2}x =  H  = \int \mathcal{H} d^{2}x$
($\mathcal{H}$ is the density of the Hamiltonian $H$) at a fixed  value  of the
Noether  charge $Q_{\chi} = \int j_{\chi }^{0}d^{2}x$.
In  this  case,  the  soliton  solution  is  an  unconditional  extremum of the
functional
\begin{equation}
F=\int \mathcal{H}d^{2}x-\omega \int j_{\chi }^{0}d^{2}x =
H - \omega Q_{\chi},                                                 \label{14}
\end{equation}
where $\omega$ is the Lagrange multiplier.
The Noether charge $Q_{\chi}$ is written in terms of the canonically conjugated
fields  $\chi$,  $\chi^{\ast}$,  $\pi_{\chi }  =  \partial \mathcal{L}/\partial
\left(\partial_{0} \chi \right) = \left(D_{0} \chi \right)^{\ast}$,  and $\pi_{
\chi ^{\ast }}= \partial \mathcal{L}/\partial  \left( \partial _{0} \chi^{\ast}
\right)  = D_{0} \chi$ as follows:
\begin{equation}
Q_{\chi } = - i \int \left( \chi \pi _{\chi }-\chi ^{\ast }\pi _{\chi ^{\ast
}}\right) d^{2}x.                                                    \label{15}
\end{equation}
From Eq.~(\ref{15}), we  obtain  the variation of the Noether charge $Q_{\chi}$
in terms of the canonically conjugate fields:
\begin{equation}
\delta Q_{\chi } = - i \int \left( \chi \delta \pi_{\chi } + \pi_{\chi } \delta
\chi - \text{c.c.}\right) d^{2}x.                                     \label{16}
\end{equation}
At the same time,  the  first  variation  of the functional $F$ vanishes for the
soliton  solution:
\begin{equation}
\delta F = \delta H - \omega \delta Q_{\chi } = 0.                   \label{17}
\end{equation}
From Eqs.~(\ref{16}) and (\ref{17}),  we obtain  the  following  Hamilton field
equations:
\begin{equation}
\partial _{0}\chi = \frac{\delta H}{\delta \pi _{\chi }} = - i \omega \chi
,\quad\partial _{0}\chi ^{\ast }=\frac{\delta H}{\delta \pi_{\chi^{\ast }}}
= i \omega \chi^{\ast},                                              \label{18}
\end{equation}
while time derivatives of the  other model's fields are equal to zero.
We see  that  the  scalar  field $\chi$ has the time dependence of Q-ball type:
\begin{equation}
\chi \left( x\right) = \chi \left( \mathbf{x}\right)
\exp \left( - i \omega t \right),                                    \label{19}
\end{equation}
whereas  the other model's fields do  not  depend on time for the adopted gauge
condition $\partial_{0} \phi = 0$.
Extremum condition (\ref{17}) leads to the important  relation  for the soliton
solution:
\begin{equation}
\frac{dE}{dQ_{\chi }} = \omega,                                      \label{20}
\end{equation}
where it is  understood  that the Lagrange multiplier $\omega$ is some function
of the Noether charge $Q_{\chi}$.

\section{The ansatz and some properties of the solution}
\label{seq:III}

To solve  field  equations  (\ref{8a}) -- (\ref{8c}), we  apply  the  following
ansatz:
\begin{align}
A^{\mu }\left( x\right) &= \left( \frac{A_{0}\left( r\right) }{e r},\frac{1}
{e r} \epsilon_{ij} n_{j} A\left( r\right) \right) , \nonumber \\
\phi \left( x\right)  & = v\exp \left( i N \theta \right) F\left( r\right),
\nonumber \\
\chi \left( x\right) & = \sigma \left( r\right) \exp \left(-i\omega t\right),
                                                                     \label{21}
\end{align}
where $\epsilon_{ij}$  are  the components of the two-dimensional antisymmetric
tensor $\left( \epsilon_{12}  =  1  \right)$  and  $n_{j}$  are  those  of  the
two-dimensional radial unit  vector $\mathbf{n} = \left(\cos\left(\theta\right)
\!,\,\sin\left(\theta\right)\right)$.
We suppose  that  the  function  $\sigma\left( r \right)$  is  real, so  ansatz
(\ref{21}) completely fixes the model's gauge.

It can  be  shown  that  ansatz  (\ref{21})  is consistent with field equations
(\ref{8a}) -- (\ref{8c}),  so  we  get  the  system  of second order  nonlinear
differential equations for the ansatz functions:
\begin{eqnarray}
&& A_{0}^{\prime \prime }(r)-\frac{A_{0}^{\prime }(r)}{r}+\frac{A_{0}(r)}
{r^{2}} - \mu A^{\prime }(r) \nonumber \\
&& - 2 \left(e^{2}v^{2}F\left( r\right) ^{2} + q^{2}\sigma \left( r\right)
^{2}\right)  \nonumber \\
&& \times A_{0}(r) + 2 e q \omega r \sigma \left( r \right)^{2} = 0,
                                                                     \label{22}
\end{eqnarray}
\begin{eqnarray}
&& A^{\prime \prime }(r)-\frac{A^{\prime }(r)}{r}-\mu A_{0}^{\prime }(r)
\nonumber \\
&& -2e^{2}v^{2}\left( N+A(r)\right) F\left( r\right) ^{2}
\nonumber \\
&& -2q^{2}\sigma \left(r\right)^{2}A\left( r\right)+\mu\frac{A_{0}(r)}{r}=0,
                                                                     \label{23}
\end{eqnarray}
\begin{eqnarray}
&& F^{\prime \prime }(r)+\frac{F^{\prime }(r)}{r} -
\frac{F(r)}{r^{2}} \nonumber \\
&& \times \left((N + A(r))^{2}-A_{0}(r)^{2}\right) \nonumber \\
&& +\lambda v^{2}\left( 1-F(r)^{2}\right) F(r) = 0,                  \label{24}
\end{eqnarray}
\begin{eqnarray}
&& \sigma ^{\prime \prime }(r)+\frac{\sigma ^{\prime }(r)}{r}+
\sigma \left( r\right) \nonumber  \\
&& \times \left( \left(\omega - \frac{q}{e}\frac{A_{0}
\left( r\right)}{r}\right)^{2}-\frac{q^{2}}{e^{2}}
\frac{A\left( r\right) ^{2}}{r^{2}}\right)  \label{25} \\
&& -\left( m^{2}-2g\sigma \left( r\right) ^{2}+3h\sigma
\left( r\right)^{4}\right) \sigma \left( r\right)  = 0.  \nonumber
\end{eqnarray}
The expression for the energy  density $\mathcal{E}=T_{00}$ can also be written
in terms of the ansatz functions:
\begin{flalign}
\mathcal{E} =& \,\frac{1}{2}\frac{A^{\prime }{}^{2}}{e^{2}r^{2}}+\frac{1}{2}
\left( \left( \hspace{-0.03in}\frac{A_{0}}{er}\hspace{-0.03in}\right)
^{\prime }\right) ^{2}+v^{2}F^{\prime }{}^{2}  \nonumber \\
& +\frac{\left( (N + A)^{2}+A_{0}{}^{2}\right) }{r^{2}}
v^{2}F^{2}                                     \nonumber \\
&+\frac{\lambda }{2}v^{4}\left( F^{2}-1\right) ^{2}+\sigma ^{\prime }{}^{2}
                                               \nonumber \\
&+\left( \omega -q\frac{A_{0}}{er}\right) ^{2}\sigma ^{2}+\frac{q^{2}}{e^{2}
}\frac{A^{2}}{r^{2}}\sigma ^{2}                \nonumber \\
&+m^{2}\sigma ^{2}-g\sigma ^{4}+h\sigma^{6}.                         \label{26}
\end{flalign}

The regularity of  the  soliton  solution  at $r = 0$ and the finiteness of the
soliton's energy $E = 2\pi \int\nolimits_{0}^{\infty}\mathcal{E}\left(r\right)r
dr$ lead us to  the  following  boundary  conditions  for the ansatz functions:
\begin{align}
A_{0}(0) &= 0, \quad A_{0}(r) \underset{r\rightarrow \infty }{\longrightarrow}0,
\nonumber \\
A(0) &= 0, \quad \, A(r) \underset{r\rightarrow \infty }{\longrightarrow} - N,
\nonumber \\
F(0) &= 0, \quad \, F(r) \underset{r\rightarrow \infty }{\longrightarrow }1,
\nonumber \\
\sigma^{\prime }(0) &= 0, \quad \, \sigma (r) \underset{r\rightarrow \infty}
{\longrightarrow } 0.                                                \label{27}
\end{align}
From the boundary conditions  for  $A(r)$  it follows that the magnetic flux of
the vortex-Q-ball system is quantized:
\begin{equation}
\Phi = 2 \pi \int_{0}^{\infty }B\left(r\right) rdr=\frac{2\pi }{e}N, \label{28}
\end{equation}
where $B(r)=-A^{\prime }(r)/(er)$ is the magnetic field strength.
Moreover, from   Eqs.~(\ref{9b})  and  (\ref{28})  it  follows  that the  total
electric charge of the vortex-Q-ball system is also quantized:
\begin{equation}
Q = -\frac{2 \pi}{e} \mu N.                                         \label{28h}
\end{equation}

In terms  of  the  ansatz  functions,  the  $C$-transformation  is  written  as
\begin{gather}
\omega \rightarrow - \omega ,\;\ N\rightarrow - N,\;A_{0}\rightarrow - A_{0},\;
\nonumber \\
A \rightarrow -A,\;\ F \rightarrow F,\;\sigma \rightarrow \sigma.  \label{28aa}
\end{gather}
It is  easily  shown  that  Eqs.~(\ref{22}) -- (\ref{25}) are  invariant  under
transformation (\ref{28aa}) as well as energy density (\ref{26}).
According  to  Eq.~(\ref{7a}),  $C$-transformation  (\ref{28aa})  is  the  only
discrete one that leaves Eqs.~(\ref{22}) -- (\ref{25}) invariant.
All  other   discrete  transformations  ($P$,  $CP$,  and  $T$)  do  not  leave
Eqs.~(\ref{22}) -- (\ref{25}) invariant.
We see that transformation (\ref{28aa}) changes  the  sign  of $\omega$, but at
the same time, it  also  changes  the  sign  of  the  soliton's winding number.
From this it follows  that  the  energy  of a soliton with a given fixed $N$ is
not invariant under the change of sign of the phase frequency: $E \left(-\omega
\right) \neq E\left(\omega \right)$.
Similarly,  it  can  be  shown  that $Q_{\phi ,\chi}\left(-\omega\right) \neq -
Q_{\phi ,\chi }\left(\omega \right)$,  so  the  absolute  values of the Noether
charges are also not invariant.

From Eqs.~(\ref{22}) -- (\ref{25}) and boundary conditions (\ref{27}) it follows
that at $r = 0$, the  power  expansion  of  the  soliton solution has the form:
\begin{align}
A_{0}\left( r\right)& =a_{1}r+\frac{a_{3}}{3!}r^{3}+O\left( r^{5}\right),
\nonumber \\
A\left( r\right)& =\frac{b_{2}}{2!}r^{2}+\frac{b_{4}}{4!}r^{4}+
O\left(r^{6}\right),
\nonumber \\
F\left( r\right)& =\frac{c_{\left|N\right|}}{\left|N\right|!}r^{\left|N\right|}
\!+\!\frac{c_{\left|N\right|+2}}{\left( \left|N\right|\!+\!2\right)!}
r^{\left|N\right|+2}\!+\!O\!\left(\!r^{\left|N\right|+4}\!\right)\!,
\nonumber \\
\sigma \left( r\right)& =d_{0}+\frac{d_{2}}{2!}r^{2}+O\left( r^{4}\right).
\label{28a}
\end{align}
In  Eq.~(\ref{28a}),  the   expressions  of  the  next-to-leading  coefficients
$a_{3}$, $b_{4}$, $c_{\left| N \right| + 2}$, and  $d_{2}$ are
\begin{align}
a_{3} = & \, 3 q d_{0}^{2}\left( a_{1}q - e\omega \right)+\frac{3}{2}\mu b_{2},
\nonumber \\
b_{4} = & \, 3 \left( q^{2}b_{2}d_{0}^{2}+2 N e^{2}v^{2}c_{\left|N\right|}^{2}
\delta_{1,\left|N\right|}\right) + \mu a_{3}, \nonumber \\
c_{\left|N\right|+2} = & -\frac{c_{\left|N\right|}}{4}
\left(\left|N\right|+2\right)\left( a_{1}^{2}+\left|N\right|\left|b_{2}\right|
+\lambda v^{2}\right), \nonumber \\
d_{2} = & \frac{d_{0}}{2}\left[ d_{0}^{2}\left( 3d_{0}^{2}h-2g\right) \right.
\nonumber \\
&\left. +e^{-2}\left( qa_{1}+e\left( m-\omega \right) \right) \right.
\nonumber \\
&\left. \times \left( -qa_{1}+e\left( m+\omega \right) \right) \right],
\label{28b}
\end{align}
where $\delta_{1,\left|N\right|}$ is the Kronecker symbol.
Linearization   of  Eqs. (\ref{22}) -- (\ref{25})  at  large  $r$  and  use  of
corresponding boundary conditions (\ref{27}) lead  us to the asymptotic form of
the solution as $r \rightarrow \infty$:
\begin{align}
A_{0}\left( r\right)  &\sim a_{\infty }\sqrt{m_{A}r}\exp \left(
-m_{A}r\right), \nonumber \\
A\left( r\right)  &\sim - N + a_{\infty }\sqrt{m_{A}r}\exp \left(
-m_{A}r\right), \nonumber \\
F\left( r\right)  &\sim 1 + c_{\infty}\frac{\exp \left(-m_{\phi }r\right)}
{\sqrt{m_{\phi }r}}, \nonumber \\
\sigma \left( r\right)&  \sim  d_{\infty }\frac{\exp
\left(-\sqrt{m^{2}-\omega^{2}} r \right) }
{\sqrt{m r}},                                                       \label{28c}
\end{align}
where
\begin{equation}
m_{A}=\sqrt{2e^{2}v^{2}+\frac{\mu ^{2}}{4}} - \frac{\mu }{2}        \label{28d}
\end{equation}
is the mass of the gauge boson  and $m_{\phi} = \sqrt{2 \lambda} v$ is the mass
of the scalar $\phi$-particle.

For symmetric  energy-momentum  tensor (\ref{10}),  the angular momentum tensor
has the form
\begin{equation}
J^{\lambda \mu \nu }=x^{\mu }T^{\lambda \nu }-x^{\nu }T^{\lambda \mu }.
                                                                     \label{29}
\end{equation}
Use  of   Eqs.~(\ref{10}), (\ref{21}),  and  (\ref{29}) leads us to the angular
momentum's density expressed in terms of the ansatz functions:
\begin{align}
\mathcal{J} &=\frac{1}{2}\epsilon _{ij}J^{0ij}=- r B E_{r}+2\frac{q}{e}A\left(
\omega - q\frac{A_{0}}{er}\right) \sigma ^{2} \nonumber \\
&-2\frac{A_{0}\left( N+A\right) }{r}v^{2}F^{2}.                      \label{30}
\end{align}
In Eq.~(\ref{30}), $E_{r}(r)=-\left( A_{0}\left(r\right)/\left( er\right)\right
)^{\prime}$ is the radial electric field strength.
Integrating the term $-rBE_{r}=-e^{-2}A^{\prime }\left(A_{0}/r\right)^{\prime}$
by parts,  taking  into  account  boundary   conditions  (\ref{27}), and  using
Eq.~(\ref{22}) to eliminate $A_{0}^{\prime \prime}$, we  obtain  the expression
for the  soliton's  angular  momentum  $J = 2 \pi \int_{0}^{\infty} \mathcal{J}
\left(r\right)rdr$:
\begin{equation}
J = -4 \pi N v^{2} \int_{0}^{\infty }A_{0}\left( r\right)F^{2}(r)dr +
\pi \frac{\mu}{e^{2}}N^{2}.                                          \label{31}
\end{equation}
At the  same  time,  Eqs.~(\ref{7})  and  (\ref{21})  lead  us to the following
expression of the Noether charge $Q_{\phi}$:
\begin{equation}
Q_{\phi }=-4\pi v^{2}\int_{0}^{\infty}A_{0}\left(r\right)F^{2}(r)dr. \label{32}
\end{equation}
From  Eqs.~(\ref{9b}), (\ref{28h}), (\ref{31}),  and (\ref{32}) it follows that
for the vortex-Q-ball system  the important relation  holds between the angular
momentum $J$  and the Noether charges $Q_{\phi}$ and $Q_{\chi}$:
\begin{flalign}
& J = N Q_{\phi }+\pi \frac{\mu }{e^{2}}N^{2} = -N\frac{q}{e}Q_{\chi } -
\pi \frac{\mu }{e^{2}}N^{2}.                                         \label{33}
\end{flalign}
We see that the angular momentum depends linearly on the Noether charges of the
scalar fields.

\begin{figure}[t]
\includegraphics[width=7.8cm]{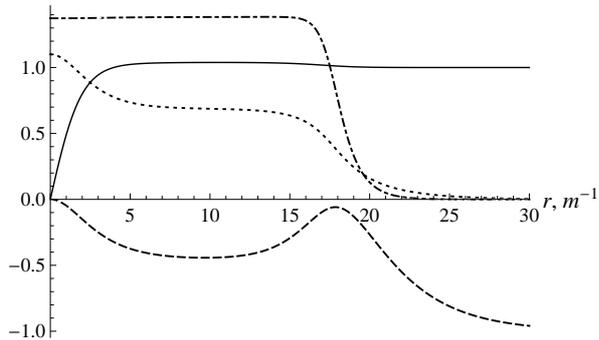}
\caption{\label{fig1}  The numerical solution $m^{-1/2}A_{0}(r)/(er)$ (dotted),
$A(r)$ (dashed), $F(r)$ (solid), and  $m^{-1/2}\sigma(r)$ (dash-dotted) for the
vortex-Q-ball system. The model's parameters are $e = q = 0.3\,m^{1/2}$, $\mu =
0.5\,  m$,  $\lambda  =  0.335\, m$, $v  =  1.221\, m^{1/2}$, $g=1.0\, m$, $h =
0.26$, and $N = 1$.  The  phase  frequency $\omega = 0.38\,m$.}
\end{figure}

Any solution of field equations (\ref{8a}) -- (\ref{8c}) is an  extremum of the
action $S=\int \mathcal{L}d^{2}xdt$.
It is readily  seen, however,  that  the  Lagrangian density (\ref{1}) does not
depend  on  time  if the  field  configurations are those of ansatz (\ref{21}).
It follows that any solution of system (\ref{22}) -- (\ref{25}) is  an extremum
of the Lagrangian $L = \int \mathcal{L} d^{2}x$.
Let $A_{0}(r)$,  $A(r)$,  $F(r)$,  and  $\sigma(r)$  be  a  solution  of system
(\ref{22}) -- (\ref{25}) satisfying boundary conditions (\ref{27}).
After the scale transformation of the solution's argument $r \rightarrow\lambda
r$, the  Lagrangian $L$  becomes  a  function of the scale parameter $\lambda$.
The function $L\left(\lambda\right)$ has an extremum at $\lambda =  1$,  so its
derivative with respect to $\lambda$ vanishes at this point: $\left. dL/d\lambda
\right\vert_{\lambda = 1} = 0$.
From  this  equation  it  follows  that  the  virial  relation  holds  for  the
vortex-Q-ball system:
\begin{equation}
2 \left( E^{\left( E\right) }-E^{\left( H\right) }+E^{\left( P\right)
}\right) +L^{\left(C\!S\right) }-\omega Q_{\chi } = 0,               \label{34}
\end{equation}
where
\begin{equation}
E^{\left( E\right) }=\frac{1}{2}\int E_{i}E_{i} d^{2}x =
\pi \int_{0}^{\infty}\left( \left( \hspace{-0.03in}
\frac{A_{0}}{er}\hspace{-0.03in}\right)^{\prime }\right)^{2}r dr     \label{35}
\end{equation}
is the electric field's energy,
\begin{equation}
E^{\left( B\right) }=\frac{1}{2}\int B^{2}d^{2}x =
\pi \int_{0}^{\infty} \frac{A^{\prime}{}^{2}}{e^{2}r} dr             \label{36}
\end{equation}
is the magnetic field's energy,
\begin{equation}
E^{\left( P\right) }=2\pi\int_{0}^{\infty}
\left[V\left( \left\vert \phi \right\vert \right)
+U\left( \left\vert \chi \right\vert \right) \right] r dr            \label{37}
\end{equation}
is the potential part of the soliton's energy, and
\begin{equation}
L^{\left(C\!S\right) }=\frac{\mu }{4}\int \epsilon ^{\rho \sigma \tau}
F_{\rho \sigma}A_{\tau}d^{2}x                                       \label{37a}
\end{equation}
is the Chern-Simons part of the model's Lagrangian.

\section{Numerical results}
\label{seq:IV}

\begin{figure}[t]
\includegraphics[width=7.8cm]{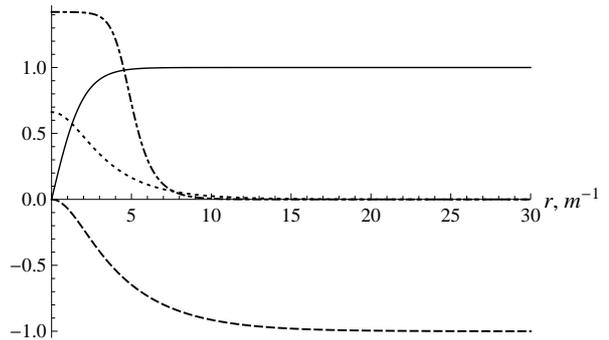}
\caption{\label{fig2}  The numerical solution $m^{-1/2}A_{0}(r)/(er)$ (dotted),
$A(r)$ (dashed), $F(r)$ (solid), and  $m^{-1/2}\sigma(r)$ (dash-dotted) for the
system of vortex  and  Q-ball that do not interact with each other. The model's
parameters are the same as in Fig.~1.}
\end{figure}

Now we  present  some  numerical  results  concerning the vortex-Q-ball system.
For numerical  calculations,  we  use  the natural units $c = 1$,  $\hbar = 1$.
In addition, the mass $m$ of scalar $\chi$-particle is used as the energy unit.
After that, the model is  completely  determined  by the seven parameters: $e$,
$q$, $\mu$, $\lambda$, $v$, $g$, and $h$.
We chose  the  following  values  of  these parameters: $e = q = 0.3\,m^{1/2}$,
$\mu=0.5\,m$, $\lambda=0.335\,m$, $v=1.221\,m^{1/2}$, $g=1.0\,m$, and $h=0.26$,
where  the  parameters'  dimensions   correspond   to   the $(2+1)$-dimensional
case.
The  correctness   of   the   numerical   solution   were   checked  by  use of
Eqs.~(\ref{9b}), (\ref{20}), (\ref{33}), and (\ref{34}).

In Fig.~1, we can see the dimensionless  zero component $m^{-1/2}A_{0}(r)/(er)$
of the gauge potential along with the  dimensionless  ansatz  functions $A(r)$,
$F(r)$, and $m^{-1/2}\sigma(r)$.
The vortex part of the soliton system is in the topological sector  with $N=1$,
the phase frequency $\omega$ is equal to $0.38\,m$.
Figure 2 presents  the  numerical  solution  for  the case $q = 0$, whereas the
other model's parameters remain the same as in Fig.~1.
This corresponds to superimposed gauged  vortex  and  non-gauged Q-ball that do
not interact with each other.
From Figs.~1 and 2, we  can  conclude  that  the  gauge interaction between the
vortex and Q-ball components leads to significant changes  in the shapes of the
ansatz functions $A_{0}(r)$, $A(r)$, and $\sigma(r)$, while the shape of $F(r)$
does not change significantly.

\begin{figure}[t]
\includegraphics[width=7.8cm]{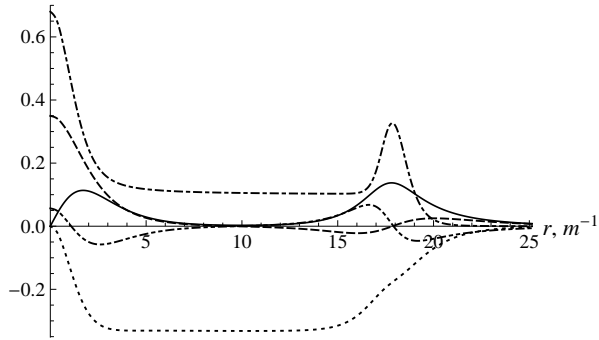}
\caption{\label{fig3} The dimensionless versions of the electric field strength
$\tilde{E}_{r}(r)  =  m^{-3/2}E_{r}(r)$ (solid), the  magnetic  field  strength
$\tilde{B}(r)  =  m^{-3/2}B(r)$  (dashed), the  scaled  energy  density  $0.5\,
\tilde{\mathcal{E}}(r) = 0.5\,m^{-3}\mathcal{E}(r)$ (dash-dotted), the electric
charge density $\tilde{j}_{0}(r) = m^{-5/2}j_{0}(r)$ (dash-dot-dotted), and the
scaled  angular  momentum's  density $0.5\,\tilde{\mathcal{J}}(r) = 0.5\,m^{-2}
\mathcal{J}(r)$ (dotted), corresponding to the solution in Fig.~1.}
\end{figure}

Figure 3 shows  the  dimensionless  versions  of  the  electric  field strength
$\tilde{E}_{r}(r)=m^{-3/2}E_{r}(r)$, the magnetic field strength $\tilde{B}(r)=
m^{-3/2}B(r)$, the scaled  energy  density $0.5\,\tilde{\mathcal{E}}(r) = 0.5\,
m^{-3}\mathcal{E}(r)$, the electric charge  density  $\tilde{j}_{0}(r)=m^{-5/2}
j_{0}(r)$, and  the scaled angular momentum's density $0.5\,\tilde{\mathcal{J}}
(r) = 0.5\,m^{-2} \mathcal{J}(r)$  that  correspond  to the soliton solution in
Fig.~1.
We see that just as in \cite{loginov_plb_777}, the vortex-Q-ball system consists
of three  parts:  the  central  transition  region, the  inner  region, and the
external transition region.
We also  see  that the  densities  of  the  energy and the angular momentum are
approximately constant in the inner region, while  the  electric  and  magnetic
field strengths are close to zero there.

In Fig.~4, we can see the  dimensionless  soliton  energy $\tilde{E} = m^{-1}E$
as a function of  the  dimensionless  phase frequency  $\tilde{\omega} = m^{-1}
\omega$.
The function $\tilde{E}\left( \tilde{\omega} \right)$ is presented in the range
from the minimum values of $\left\vert \tilde{\omega} \right\vert$ that we have
reached by numerical methods to its maximum value of $1$.
The most  important  feature  of  Fig.~4  is  that  the soliton's energy is not
invariant under  the  change  of  sign of the phase frequency: $\tilde{E}\left(
-\tilde{\omega}\right) \neq \tilde{E}\left( \tilde{\omega} \right)$.
This fact  is  a  direct  consequence  of the $T$-invariance breaking, which is
caused by the Chern-Simons term in the Lagrangian (\ref{1}).
From Fig.~4  it  follows  that  the  soliton's  energy $E$ tends to infinity as
$\left  \vert  \tilde{ \omega}  \right  \vert$  tends  to  its  minimum  values
(thin-wall regime).
In the  thin-wall  regime, the  spatial  size  of  the  soliton's  inner region
increases indefinitely, so  the  main  contribution to the soliton's energy and
angular momentum comes from this region.
\begin{figure}[t]
\includegraphics[width=7.8cm]{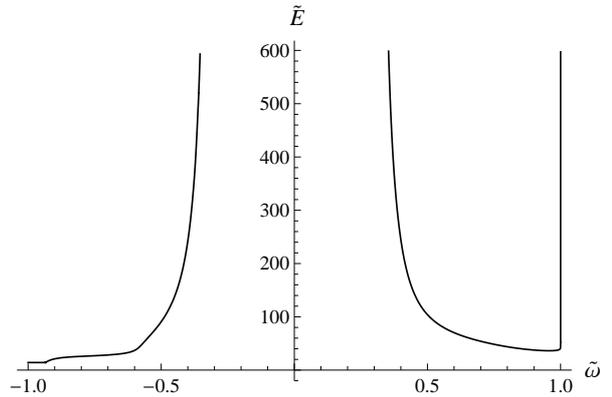}
\caption{\label{fig4} The  dimensionless  soliton  energy $\tilde{E} = m^{-1}E$
as a function of  the  dimensionless  phase  frequency $\tilde{\omega} = m^{-1}
\omega$.  The  model's  parameters are the same as in Fig.~1.}
\end{figure}

As $\tilde{ \omega} \rightarrow 1$,  the  vortex-Q-ball  system  goes  into the
thick-wall regime.
As well as in the thin-wall regime, the soliton's Noether charge $Q_{\chi}$ and
energy $E$ tend to infinity in the thick-wall regime.
It was found numerically that $Q_{\chi}(\tilde{\omega})$ and $\tilde{E}(\tilde{
\omega})$  have  the  following  behaviour  as $\tilde{ \omega} \rightarrow 1$:
\begin{eqnarray}
& & Q_{\chi }  \underset{\tilde{\omega} \rightarrow 1}{\longrightarrow}  B +
A \left(1-\tilde{\omega} \right)^{-\frac{1}{2}}, \nonumber \\
& & \tilde{E}  \underset{\tilde{\omega} \rightarrow 1}{\longrightarrow}  C +
A\left(2-\tilde{\omega}\right) \left( 1-\tilde{\omega} \right)^{-\frac{1}{2}},
                                                                     \label{38}
\end{eqnarray}
where $A$, $B$, and $C$ are positive constants.
From Eq.~(\ref{38}) it  follows  that  the  behaviour of $Q_{\chi}\left(\tilde{
\omega}\right)$ and $\tilde{E}\left(\tilde{\omega}\right)$  in  the  thick-wall
regime is in agreement with Eq.~(\ref{20}).
Such  behaviour of $Q_{\chi}(\tilde{\omega})$  and  $\tilde{E}(\tilde{\omega})$
in a neighborhood of the  maximum  value $\tilde{\omega} = 1$ is very different
from that of the  two-dimensional non-gauged Q-ball \cite{lee}.
It is also quite different from  the  behaviour  of   the  vortex-Q-ball system
\cite{loginov_plb_777} in the  Maxwell gauge model.
However, the  behaviour of  $Q_{\chi}(\tilde{\omega})$  and  $\tilde{E}(\tilde{
\omega})$ in a neighborhood of $\tilde{\omega} = 1$ is  similar  to that of the
usual three-dimensional Q-ball \cite{lee}.

In Fig.~5, we can  see  the  dependences of the  dimensionless  soliton  energy
$\tilde{E}$  and  the  absolute value  of  Noether  charge $Q_{\chi}$ (which is
negative for $\tilde{\omega} < 0$) on the absolute value of $\tilde{\omega}$ in
a neighborhood of $\left\vert \tilde{\omega} \right\vert = 1$.
From Fig.~5 it follows that the  behaviour  of  the vortex-Q-ball system in the
neighborhoods  of  $\tilde{\omega} = -1$ and $\tilde{\omega} = 1$ is completely
different.
Indeed, its  behaviour  near $\tilde{\omega}  =  1$ corresponds  to  thick-wall
regime (\ref{38}). 
At the same time, its  behaviour  near $\tilde{\omega} = -1$ is rather unusual.
Firstly, there is no thick-wall regime here.
Secondly, the  Q-ball  component  of  the  vortex-Q-ball  system  disappears at
$\tilde{\omega} = -0.93$.
For $\tilde{\omega}  \in  \left( -1, -0.93 \right)$,  we  have  only the single
Maxwell-Chern-Simons vortex without any Q-ball component.
\begin{figure}[t]
\includegraphics[width=7.8cm]{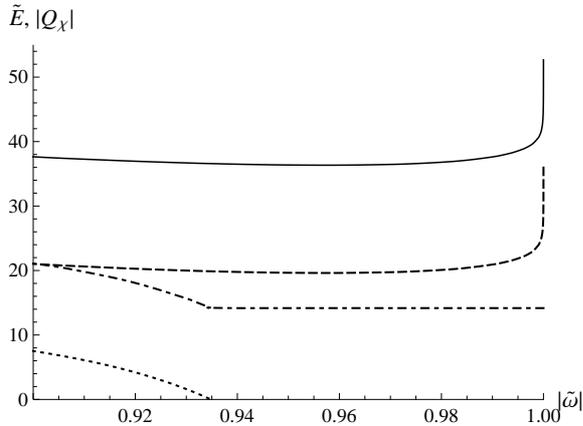}
\caption{\label{fig5}  The dimensionless  soliton  energy $\tilde{E} = m^{-1}E$
(solid for $\tilde{\omega} > 0$  and dash-dotted for $\tilde{\omega} < 0$)  and
the absolute value of Noether charge $Q_{\chi}$ (dashed for $\tilde{\omega} >0$
and dotted for $\tilde{\omega} < 0$)  as  functions  of the  absolute  value of
dimensionless  phase  frequency  $\left \vert \tilde{\omega} \right \vert$ in a
neighborhood  of  $\left\vert \tilde{\omega} \right\vert = 1$.}
\end{figure}

Figure  6  shows  the  dimensionless  energy  $\tilde{E}$  as a function of the
absolute  value   of   Noether   charge  $Q_{\chi}$   for  the  both  signs  of
$\tilde{\omega}$.
It also shows the similar dependence $\tilde{E}\left(\left\vert Q_{\chi} \right
\vert\right)$ for the two-dimensional non-gauged Q-ball with the same parameters
$m$,  $g$,  and  $h$  as  for  the  vortex-Q-ball system.
In addition, the straight line $\tilde{E} = \left\vert Q_{\chi} \right\vert$ is
also shown in Fig.~6.
We  can  see  that  the  curve $\tilde{E}( \left\vert  Q_{\chi}  \right\vert )$
corresponding to the two-dimensional Q-ball  is  tangent to  the  straight line
$\tilde{E} = \left\vert Q_{\chi} \right\vert$  at  some nonzero $\left\vert Q_{
\chi} \right\vert$ as it should be \cite{lee}.
In contrast to this, the vortex-Q-ball  system  is described by the two curves,
which correspond  to  the  both  signs of the phase frequency $\tilde{\omega}$.
The curve $\tilde{E}(Q_{\chi})$ corresponding to the  positive $\tilde{\omega}$
is similar to that of three-dimensional Q-ball.
In particular, it has the cusp and consists of two branches.
As $Q_{\chi} \rightarrow \infty$,  the  lower  branch  goes  into the thin-wall
regime, while the upper one goes into the thick-wall regime.
At the same time, the curve $\tilde{E}(-Q_{\chi})$ corresponding to the negative
$\tilde{\omega}$ has no cusp and consists of only one branch.
The curve  starts  at $Q_{\chi} = 0$  and  goes  into  the  thin-wall regime as
$Q_{\chi} \rightarrow -\infty$.
From Fig.~6, we can conclude that  in the thin-wall regime the Q-ball component
of the the vortex-Q-ball system is stable to the decay into the  massive scalar
$\chi$-particles.
\begin{figure}[t]
\includegraphics[width=7.8cm]{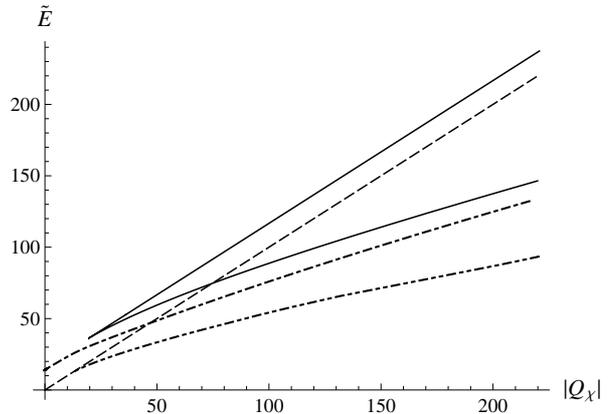}
\caption{\label{fig6} The vortex-Q-ball system's dimensionless energy $\tilde{E
}$  as  a  function  of  the  absolute  value  of Noether charge $Q_{\chi}$ for
$\tilde{\omega} > 0$ (solid) and for $\tilde{\omega}  <  0$  (dash-dotted), and
that for the two-dimensional non-gauged  Q-ball (dash-dot-dotted) with the same
parameters $m$, $g$, and $h$  as  for the vortex-Q-ball system. The dashed line
is the straight line $\tilde{E} = \left\vert Q_{\chi} \right\vert$.}
\end{figure}

\section{Conclusions}
\label{seq:V}

In the present paper,  we  have  researched  the soliton system consisting of a
vortex and a Q-ball that  interact  with  each  other  through a common Abelian
gauge field.
Like a  vortex, this  two-dimensional  soliton  system  has  quantized magnetic
flux (\ref{28}).
Due to the Chern-Simons  term  in  the  Lagrangian, the quantized magnetic flux
leads to quantized electric  charge (\ref{28h}) of the soliton system and, as a
consequence, to a nonzero radial electric field.
As a result, the soliton  system  possesses nonzero angular momentum (\ref{33})
that depends linearly on the Noether charges of the scalar fields.
Owing to the Chern-Simons  term,  the energy of the vortex-Q-ball system is not
invariant under the sign reversal of the phase frequency $\omega$.
This in turn leads to the significant change of the dependence $E(Q_{\chi})$ in
comparison with  the  vortex-Q-ball  system \cite{loginov_plb_777} and with the
two-dimensional non-gauged Q-ball \cite{lee}.
The   vortex-Q-ball   system   combines   properties   of  both  nontopological
(Eq.~(\ref{20}))  and  topological (boundary condition  (\ref{27})  for  $A(r)$
and,  as  a  consequence,  magnetic   flux  quantization  (\ref{28})) solitons.

Finally,  let  us  point  out  a  possible  application of the results obtained
in \cite{loginov_plb_777} and in the present paper.
A vortex-Q-ball string may arise when a cosmic string  passes through a charged
scalar condensate.
Such a condensate could exist in the early universe; electrically charged boson
stars \cite{jetzer_227}, if they  exist, also  consist of  such  a  condensate.
A part of  the  condensate  may  be  carried  away by the passing cosmic vortex
string, with the result that the vortex-Q-ball string arises.
In this case, the gauge interaction between vortex and Q-ball components of the
vortex-Q-ball  string   leads  to  significant  changes  of  their  properties.

\section*{Acknowledgments}

The research  is  carried  out  at  Tomsk  Polytechnic  University  within  the
framework of  Tomsk  Polytechnic University Competitiveness Enhancement Program
grant.





\bibliographystyle{elsarticle-num}

\bibliography{article}






\end{document}